\documentclass[12pt]{article}

\usepackage{float}
\usepackage{scicite}
\usepackage{graphicx}
\newcommand{\be}{\begin{equation}}
\newcommand{\ee}{\end{equation}}
\newcommand{\htd}{$\rm H_2D^+$}

\newcommand{\dth}{$\rm{\left[D/H\right]}$}
\newcommand{\doh}{$\rm{D/H}$}
\newcommand{\dwat}{$\rm{\left[D/H\right]_{H_2O}}$}

\newcommand\ssr{{Space~Sci.~Rev.}}%
\newcommand\icarus{{Icarus}}%
 \newcommand\apss{{Ap\&SS}}%
\newcommand\aap{{A\&A}}%
\newcommand\apj{{ApJ}}
\newcommand\apjl{{ApJL}}
\newcommand\aj{{AJ}}
\newcommand\araa{{ARA\&A}}
\newcommand\apjs{{ApJS}}
\newcommand\mnras{{MNRAS}}
  \newcommand\jcp{{J.~Chem.~Phys.}}%
  \newcommand\planss{{Planet.~Space~Sci.}}%
  \newcommand\gca{{Geochim.~Cosmochim.~Acta}}%
    \newcommand\nat{{Nature}}%
    
\usepackage{times}
\usepackage{amssymb,amsmath}
\usepackage{wasysym}

\newenvironment{sciabstract}{%
\begin{quote} \bf}
{\end{quote}}

\newcounter{lastnote}

\title{The Ancient Heritage of Water Ice in the Solar System}

\author
{L. Ilsedore Cleeves,$^{1\ast}$ Edwin A. Bergin$^{1}$, Conel M. O'D. Alexander$^{2}$, \\ Fujun Du$^{1}$, Dawn Graninger$^{3}$, Karin I. {\"O}berg$^{3}$ and Tim J. Harries$^{4}$\\
\\
\normalsize{$^{1}$Department of Astronomy, University of Michigan,}\\
\normalsize{500 Church St, MI 48109, USA}\\
\normalsize{$^{2}$DTM, Carnegie Institution of Washington,}\\
\normalsize{Washington, DC 20015, USA}\\
\normalsize{$^{3}$Harvard-Smithsonian Center for Astrophysics, Harvard University,}\\
\normalsize{Cambridge, MA 02138, USA}\\
\normalsize{$^{4}$Department of Physics and Astronomy, University of Exeter,} \\ 
\normalsize{Stocker Road, Exeter, EX4 4QL, UK}\\
\normalsize{$^\ast$To whom correspondence should be addressed; E-mail:  cleeves@umich.edu.}
}

\date{}

\begin{document} 

\baselineskip24pt

\maketitle 

\begin{sciabstract}
Identifying the source of Earth's water is central to understanding the origins of life-fostering environments and to assessing the prevalence of such environments in space.  Water throughout the solar system exhibits deuterium-to-hydrogen enrichments, a fossil relic of low-temperature, ion-derived chemistry within either (i) the parent molecular cloud or (ii) the solar nebula protoplanetary disk. Utilizing a comprehensive treatment of disk ionization, we find that ion-driven deuterium pathways are inefficient, curtailing the disk's deuterated water formation and its viability as the sole source for the solar system's water.  This finding implies that if the solar system's formation was typical, abundant interstellar ices are available to all nascent planetary systems.
\end{sciabstract}

Water is ubiquitous across the solar system,  in cometary ices, terrestrial oceans, the icy moons of the giant planets, and in the shadowed basins of Mercury \cite{encrenaz2008,lawrence2013}.  Water has left its mark in hydrated minerals in meteorites, in lunar basalts \cite{barnes2014} and in martian melt-inclusions \cite{usui2012}.  The presence of liquid water facilitated the emergence of life on Earth, and thus understanding the origin(s) of water throughout the solar system is a key goal of astrobiology.  Comets and asteroids (traced by meteorites) remain the most primitive objects, providing a natural ``time capsule'' of the conditions present during the epoch of planet formation.  Their compositions reflect those of the gas, dust, and -- most importantly -- ices encircling the Sun at its birth, i.e., the solar nebula protoplanetary disk.  There remains an open question, however, as to when and where these ices formed, whether they  i) originated in the dense interstellar medium (ISM) in the cold molecular cloud core prior to the Sun's formation, or ii) are products of reprocessing within the solar nebula \cite{robert2000,ceccarelli2014,vandishoeck2014}. Scenario i) would imply that abundant interstellar ices, including water and presolar organic material, are incorporated into all planet-forming disks.  By contrast, local formation within the solar nebula in scenario ii) would potentially result in large water abundance variations from stellar system to system, dependent upon the properties of the star and disk.

In this work, we aim to constrain the formation environment of the solar system's water using deuterium fractionation as our chemical tracer.  Water is  enriched in deuterium relative to hydrogen (\doh) compared to the initial bulk solar composition across a wide range of solar system bodies,  including comets, \cite{bockeleemorvan1998,eberhardt1995}, terrestrial and ancient Martian water \cite{usui2012}, and hydrated minerals in meteorites \cite{alexander2012}.  The amount of deuterium relative to hydrogen of a molecule depends on its formation environment, and thus the \doh\ fraction in water, \dwat, can be used to differentiate between the proposed source environments.  Interstellar ices, as revealed by sublimation in close proximity to forming young stars, also exhibit high degrees of  deuterium-enrichment, $\sim 2-30\times$ that of terrestrial water \cite{dartois2003,coutens2012,persson2012,persson2014}. It is unknown to what extent these extremely deuterated, interstellar ices are incorporated into planetesimals or if instead the interstellar chemical record is erased by reprocessing during the formation of the disk \cite{visser2009,yang2013}.   Owing to its high binding energy to grain surfaces, theoretical models predict that water is delivered from the dense molecular cloud to the disk primarily as ice with some fraction sublimated at the accretion shock in the inner tens of astronomical units (AU) \cite{visser2009}.  If a substantial fraction of the interstellar water is thermally reprocessed, the interstellar deuterated record could then be erased.  In this instance, the disk is left as primary source for (re-)creating the deuterium-enriched water present throughout our solar system.

The key ingredients necessary to form water with high \doh\ are cold temperatures, oxygen, and a molecular hydrogen ($\rm{H_2}$) ionization source.  The two primary chemical pathways for making deuterated water  are: (i) gas-phase ion-neutral reactions primarily through \htd, and (ii) grain-surface formation (ices) from ionization-generated hydrogen and deuterium atoms from $\rm{H_2}$.  Both reaction pathways depend critically on the formation of \htd.  In particular, the gas-phase channel (i) involves the reaction of \htd\ ions with atomic oxygen or OH through a sequence of steps to form $\rm{H_2DO^+}$, which recombines to form a water molecule.  The grain-surface channel (ii) is powered by \htd\ recombination with electrons or grains, which liberates hydrogen and deuterium atoms that react with oxygen atoms on cold dust grains.  \htd\ becomes enriched relative to ${\rm{H_3^{+}}}$ due to the deuterated-isotopologue being energetically favored at low temperatures.  There is an energy barrier $\Delta{E_1}$ to return to ${\rm{H_3^{+}}}$, i.e., ${\rm{H_3^{+}+HD\rightleftharpoons{H_2D^{+}}+H_{2}}}+\Delta{E_1}$, where $\Delta{E_1}\sim124$~K, though the precise value depends upon the nuclear spin of the reactants and products \cite{supmat}.  The relatively modest value of $\Delta{E_1}$ restricts deuterium enrichments in H$_3$$^+$ to the coldest gas, $T\lesssim50$~K.  Thus, deuterium-enriched water formation requires the right mix of environmental conditions: cold gas, gas-phase oxygen and ionization.
 
 The conditions in the dense interstellar medium, i.e., the cloud core, readily satisfy these requirements, where temperatures are typically $T\sim10$~K and ionization is provided via galactic cosmic rays (GCRs). In this regard, the conditions in the core and in the outermost regions of the solar nebula are often thought of as analogous \cite{aikawa1999}. This is because the outskirts of protoplanetary disks typically contain the coldest ($T\lesssim30$~K), lowest density gas, and are often assumed to be fully permeated by GCRs.  However, the efficacy of GCRs as ionizing sources in protoplanetary disks has been called into question due to the deflection of GCRs by the stellar winds produced by young stars \cite{cleeves2013a}.  Even the mild, modern-day solar wind reduces the GCR ionization rate, $\zeta_{\rm{GCR}}$, by a factor of $\sim100$ below that of the unshielded ISM.  
Limits on protoplanetary disks' molecular ion emission indicate low GCR ionization rates, $\zeta_{\rm{GCR}}\le3\times10^{-17}$~s$^{-1}$ \cite{chapillon2011}.   In the absence of GCRs, disk midplane ionization is instead provided by scattered X-ray photons  from the central star and the decay of short-lived radionuclides (SLRs), where the latter's influence decreases with time \cite{umebayashi2013}.  In addition, the core-disk analogy breaks down with regard to gas density.  At the outermost radius of the protosolar gaseous disk, $R_{\rm{out}}\sim50-80$~AU  \cite{supmat}, the typical disk density is $n\sim10^{10}$~cm$^{-3}$. This value is approximately five orders-of-magnitude higher than typical values within the interstellar molecular core \cite{minamidani2011}.  The steady state ion abundance is proportional to $\sqrt{\zeta/n}$, where $\zeta$ is the ionization rate and $n$ is the volumetric gas (hydrogen) density.  For a constant ionization rate, a density increase of $10^{5}$ corresponds to $\chi_i$ being a factor of $\sim300$ below that expected in the ISM.  Thus, wind or magnetically driven deflection of ionizing GCRs, coupled with high gas densities will strongly inhibit the disk from generating deuterium enrichments through the standard cold chemical reactions (i) and (ii), described above.
 
To test the disk hypothesis, we explore whether ionization-driven chemistry within the disk alone is capable of producing the deuterium-enriched water that was present in the early solar system.  We have constructed a comprehensive model of disk ionization, including detailed radiative transfer, reduced GCR ionization and SLR decay.  To simulate the ``reset'' scenario, i.e., all interstellar deuterium enhancement is initially lost, we start with unenriched water with bulk solar \doh\ composition, $\rm{\left[D/H\right]_{H_2}}=2.0\pm0.35\times10^{-5}$ \cite{geiss2003}, and quantify the maximum amount of deuterated water produced by chemical processes in a static protoplanetary disk over 1~Myr of evolution.  The goal is to determine whether or not the conditions present in the solar nebula were capable of producing at minimum the measured isotopically enriched water in meteorites, ocean water (VSMOW), and comets (see Fig.~1). We do not attempt to address the eventual fate of this water by additional processing, i.e., by radial or vertical mixing, which tends to reduce the bulk \doh\ ratio in water \cite{willacy2009,albertsson2014}.

Instead, our emphasis is on the physical mechanism necessary for \doh\ enrichment: ionization.   We illustrate the suite of ionization processes considered in the traditional picture of disk ionization (Fig.~2a) alongside a schematic for our new model (Fig.~2b).  More specifically, we include ``solar-maximum'' levels of GCR wind-modulation for the incident GCR rate, a factor of $\sim300$ below that of the ISM, Monte Carlo propagation of X-ray photoabsorption and scattering, and ionization by SLR decay products, including losses in the low density ($\Sigma_{\rm{gas}}\lesssim10$~g~cm$^{-2}$) regions of the disk \cite{supmat}.  The total $\rm H_2$ ionization rate is shown in Figures~2c and 2d.  It can be seen that while the warm surface layers ($\Sigma_{\rm{gas}}\lesssim1$~g~cm$^{-2}$) are strongly ionized by stellar X-rays, $\zeta_{\rm{XR}}\gtrsim10^{-15}$~s$^{-1}$, the midplane is comparatively devoid of ionization due to the modulation of  incident GCRs (Fig.~2d).

To compute molecular abundances, we compile a simplified deuterium reaction network designed to robustly predict the \doh\ in water  resulting from gas-phase and grain-surface chemistry \cite{supmat}.   We include both ion-neutral and the hot-phase neutral-neutral water chemistry \cite{bergin1995}, as well as self-shielding by HD and D$_2$, in addition to the standard chemical network \cite{supmat}. We include an updated treatment of the inherently surface-dependent CO freeze-out process motivated by new laboratory data on CO ice binding energies \cite{supmat}.  CO freeze-out is important for the present study as CO regulates the amount of oxygen present in gas above $T>17$~K and available for new water formation.   We place the majority of volatile carbon in CO and the rest of the oxygen not in CO in water ice \cite{visser2009} (see Supplementary Materials for additional runs \cite{supmat}).  We examine the final \doh\ of water ice after 1~Myr of chemical evolution, i.e., the approximate lifetime of the gas-rich disk and, correspondingly, the duration over which the disk is able to build up deuterated water (see Fig.~3).  The ions are most abundant in the upper, X-ray-dominated surface layers, whose temperatures are too warm for significant deuterium enrichment in $\rm{\left[D/H\right]_{H_3^+}}$.  The bulk ice mass is closest to the cold midplane, where only a meager amount (per unit volume) of ${\rm{H_3^{+}}}$ and ${\rm{H_2D^{+}}}$ remain in the gas, a consequence of low ionization rates and high densities.

In addition to spatial abundances, we provide ratios of the vertically integrated column densities of both ions and ices (Fig.~3, bottom).  The choice of VSMOW as a benchmark is somewhat arbitrary considering that comets exceed $\rm{\left[D/H\right]}$ in VSMOW by factors of 1-3$\times$ and meteorites have typically lower values, a factor $\sim2$ less (Fig.~1).  The column-derived $\rm{\left[D/H\right]_{H_3^+}}$ approaches -- but does not reach -- VSMOW after 1~Myr at the outer edge of the disk.  Moreover, most of the molecules that contribute to the column density ratio of $\rm{\left[D/H\right]_{H_3^+}}$ arise in an intermediate layer of cold ($T_{\rm gas}\sim30-40$~K) gas where  X-ray photons are still present. However, it is readily apparent that this enrichment does not translate into $\rm{\left[D/H\right]_{H_2O}}$.  The most deuterium-enriched water, $\rm{\left[D/H\right]_{H_2O}}=3\times10^{-5}$, is co-located with the enriched layer of $\rm{\left[D/H\right]_{H_3^+}}$; however, the water is considerably less enriched than the ions. Over the lifetime of the disk, the gas-phase and grain-surface chemical pathways do not produce deuterated water ices in significant quantities.  The low water production is due to a combination of (i)  a lack of sufficient ionization to maintain significant amounts of \htd\ in the gas, and (ii)  a lack of  atomic oxygen in the gas, locked up in ices.  We do find a super-deuterated layer of water at the disk surface ($\rm{\left[D/H\right]_{H_2O}}=5\times10^{-3}$). This layer is a direct consequence of selective self-shielding of HD relative to H$_2$, which leads to an over abundance of atomic D relative to H. Selective self-shielding is also the cause of $\rm{ \left[D/H\right]_{H_3^+}}$ falling below the initial bulk gas value inside of $R<40$~AU.  This layer does not, however, contribute significantly to the vertically integrated \dth$_{\rm{H_2O}}$ (bottom panels in Fig.~3).  

In summary, using a detailed physical ionization model, updated treatment of oxygen-bearing (CO) ice chemistry, and a simplified deuterium chemical network, we find that chemical processes in disks are not efficient at producing significant levels of highly deuterated water.  Our model predicts that disk chemistry can only produce a volume-integrated $\rm{\left[D/H\right]_{H_2O}}\lesssim2.1\times10^{-5}$, which is only slightly enriched from the bulk solar value ($2\times10^{-5}$). In terms of column density, we find that even in the most radially distant (coldest) regions water only attains $\rm{\left[D/H\right]_{H_2O}}\lesssim2.5\times10^{-5}$.  This finding implies that ion-chemistry within the disk cannot create the deuterium-enriched water present during the epoch of planet formation.  When we begin our models with interstellar $\rm{\left[D/H\right]_{H_2O}}\sim10^{-3}$, however, it is hard to ``erase'' \doh\ ratios with low-temperature disk chemistry alone.  A number of studies have invoked turbulent mixing of gas in the radial and/or vertical directions, which can reduce inner-disk deuterium enrichments in water \cite{willacy2009,furuya2013,albertsson2014}.  Moreover, a common feature of such models is that they begin with high levels of deuterated water, as high as $\rm{\left[D/H\right]_{H_2O}}\sim10^{-2}$.  In general, mixing in the vertical direction transports highly deuterium-enriched ices from the shielded midplane into the X-ray and UV irradiated warm surface layers \cite{ciesla2012,ciesla2014} where they can be reprocessed to lower \doh.  Ices transported radially inward, either entrained by gas accretion or subjected to radial migration, evaporate in the warm and dense inner disk and isotopically reequilibrate with H$_2$ or ion-neutral chemistry in hot ($T>100$~K) gas.    With our updated disk ionization model, we can now exclude chemical processes within the disk as an enrichment source term and conclude that  the solar nebula accreted and retained some amount of pristine interstellar ices. One potential explanation is that during the formation of the disk, there was an early high temperature episode followed by continued infall from deuterium-enriched interstellar ices \cite{yang2013}.    If we ignore the negligible contribution to deuterated water formation from the disk,  we can estimate the fraction of water in a particular solar system body, ${\rm X}$, that is pre-solar, $f_{\rm ISM}=\left({\rm{D/H_{X}}}-{\rm{D/H_{\odot}}}\right)/\left({\rm{D/H_{ISM}}}-{\rm{D/H_{\odot}}}\right)$, where ${\rm{D/H_{X}}}$ refers to \dwat\ in ${\rm X}$ and ${\rm{D/H_{\odot}}}=2\times10^{-5}$.  Water in the ISM ranges from a limit of \dwat$<2\times10^{-3}$ in interstellar ice \cite{dartois2003} to \dwat$=(3-5)\times10^{-4}$ for low-mass protostars, i.e., analogs to the Sun's formation environment \cite{persson2014}.  If the former, higher value reflects the ices accreted by the solar nebular disk, then at the very least, terrestrial oceans and comets should contain $ \gtrsim7\%$ and $\gtrsim14\%$ interstellar water, respectively.  If the low-mass protostellar values are representative,  the numbers become $30-50\%$ for terrestrial oceans and $60-100\%$ for comets.   Thus a significant fraction of the solar system's water predates the Sun.  These findings imply that some amount of interstellar ice survived the formation of the solar system and was incorporated into planetesimal bodies.     Consequently, if the formation of the solar nebula was typical, our work implies that interstellar ices from the parent molecular cloud core, including the most fundamental life-fostering ingredient, water, are widely available to all young planetary systems.

\bibliographystyle{Science}

\newpage



\begin{figure}[H]
  \centering
    \includegraphics[width=0.75\textwidth]{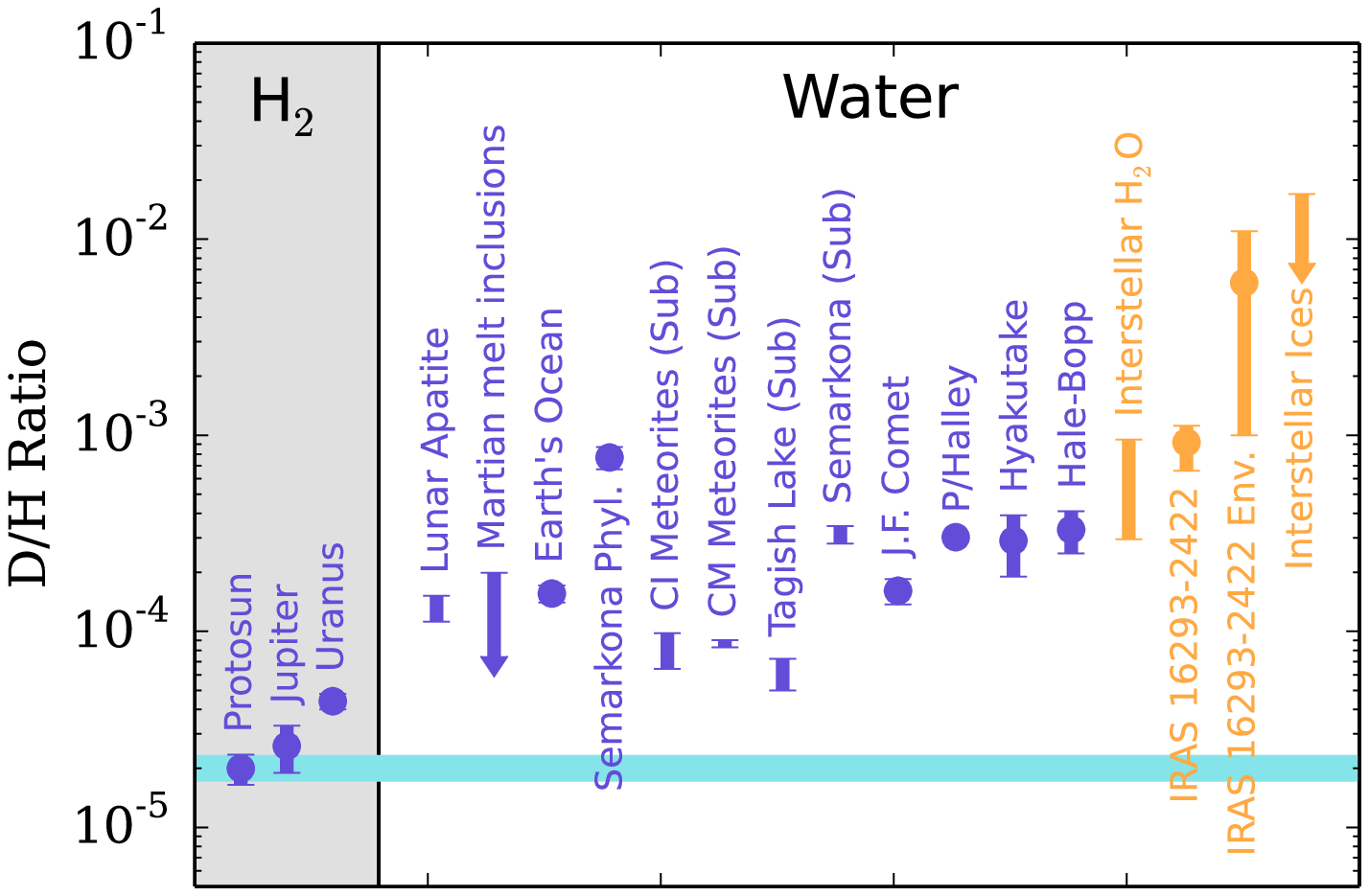} 
  \caption{{\bf Atomic D/H ratios in solar system ({\em purple}) and interstellar ({\em yellow}) sources separated into  bulk H$_2$ and water.}  Points indicate single measurements, bars without points are ranges over multiple measurements, and arrows correspond to limits.  \doh\ in the bulk gas (i.e., solar) is indicated as a horizontal blue bar.  References are provided in the online supplementary materials in Table~S4.
  }
\end{figure}

\newpage

\begin{figure}[H]
  \centering
    \includegraphics[width=1.0\textwidth]{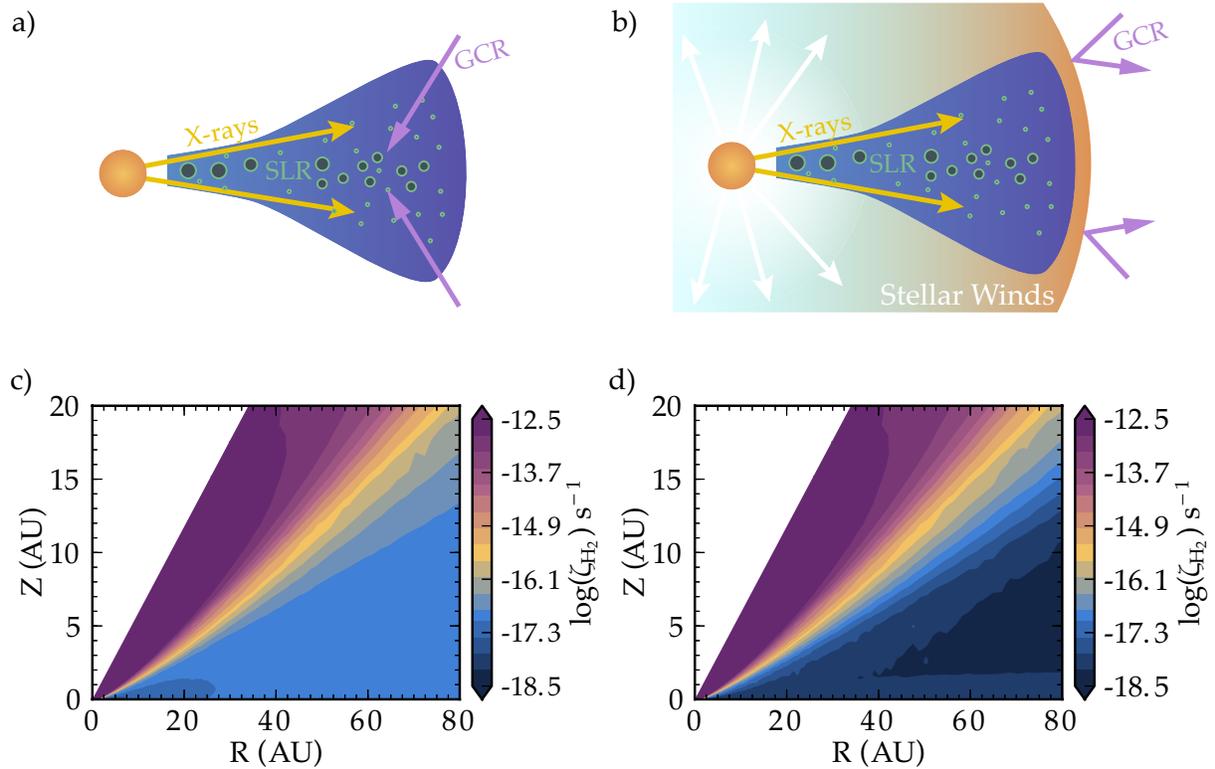}
  \caption{{\bf Schematic of energetic disk ionization sources ({\em top row}) and the calculated total H$_2$ ionization rate ({\em bottom row}). }    Panels are: {\bf (a)} a ``standard'' disk ionization model driven by X-rays and GCRs; {\bf (b)} ionization conditions under the influence of a Sun-like wind at solar maximum, now dominated by X-rays in the surface and SLRs in the midplane; {\bf (c)} calculated H$_2$ ionization rates for the scenario depicted in Panel (a); and  {\bf(d)} calculated H$_2$ ionization rates from the scenario depicted in Panel (b) and that used within the present paper.
  }
\end{figure}

\newpage

\begin{figure}[H]
  \centering
    \includegraphics[width=1.0\textwidth]{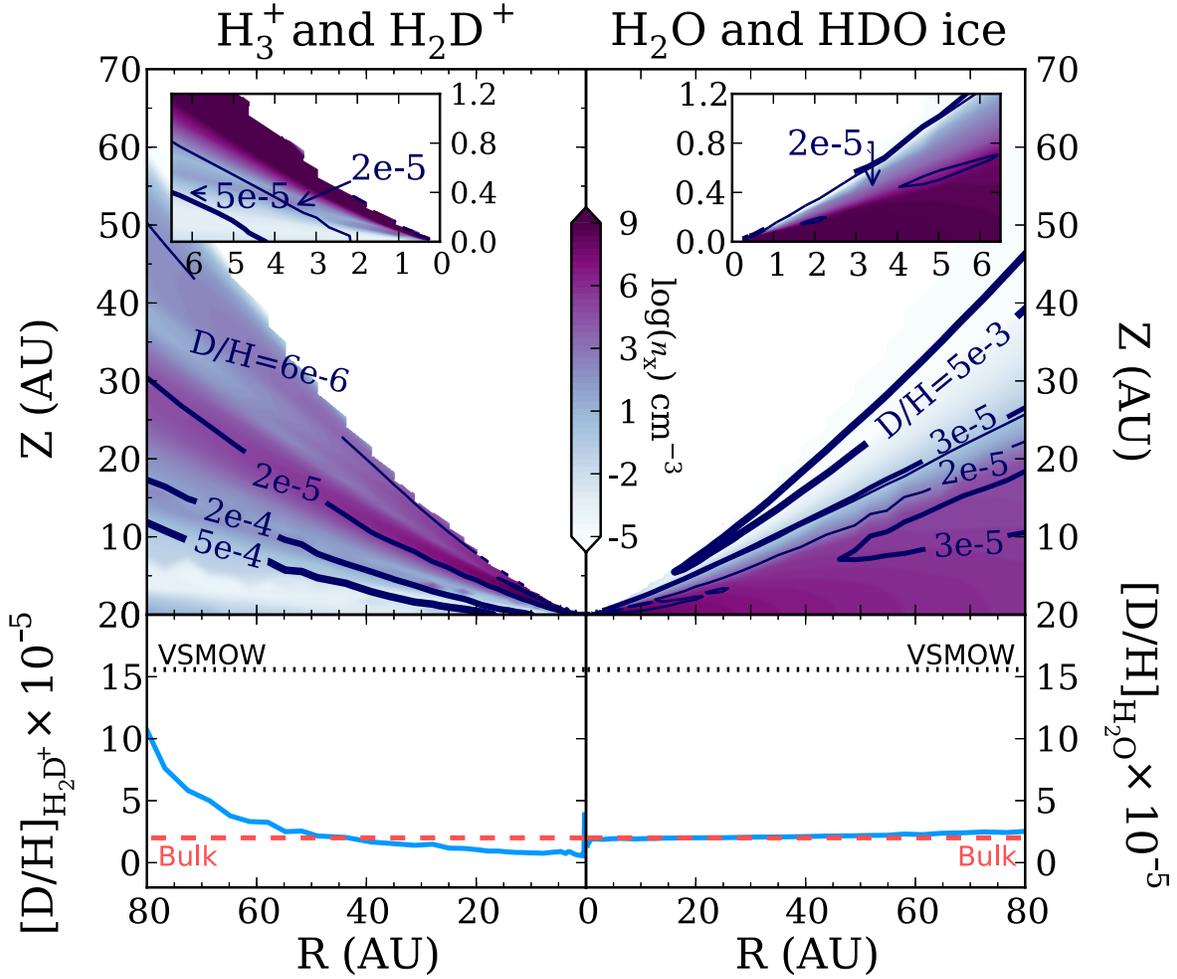}
  \caption{ {\bf Chemical abundances and column densities for the drivers of the deuterium enrichment, the ions; and the corresponding products, the ices.} Volume densities (cm$^{-3}$) of all labeled isotopologues for H$_3^+$ (left) and water (right) in filled purple contours. Solid contour lines indicate local \doh\ and are as labeled. The color scale applies to both the left and right halves of the plot.  Insets zoom-in on the inner disk with the same axes units.  Bottom panel plots show the vertically integrated \doh\ ratio of column densities versus radius. The bulk gas (protosolar) value is labeled by the red dashed line, and the Earth's ocean value (VSMOW) is labeled with the black dotted line.
  }
\end{figure}

\newpage

\noindent {\bf Acknowledgments:} LIC and EAB acknowledge support by NSF grant AST-1008800 and the Rackham Predoctoral Fellowship (LIC).  CA was partially 
supported by NASA Astrobiology grant NNA09DA81A and by NASA Cosmochemistry grant NNX11AG67G. FD was supported by NASA grant NNX12A193G.  TJH was supported by STFC grant ST/J001627/1.

\bigskip

\noindent {\bf  Supplementary Materials}

\noindent Materials and Methods

\noindent Figures S1-S4

\noindent Tables S1-S4

\noindent References (30-71)

\newpage

\section*{Supplementary Materials}
\subsection{Materials and Methods}

\noindent \begin{center} {\bf Physical Model} \end{center}

The disk physical structure is based upon a previously published model of a dust-settled disk \cite{cleeves2013a} with typical geometric and settling structural parameters reflecting a sample of modeled dust continuum observations \cite{andrews2011}.  The model has been truncated from its original outer radius of $R=400$~AU to have an outer radius of $R=80$~AU, which reflects the maximal radius of the protosolar disk as determined from dynamical models of the lack of planetesimals on high inclination orbits \cite{kretke2012}.  This corresponds to a modified disk gas mass of $M_{\rm{gas}}=0.008$~M$_{\odot}$ from its original value of $M_{\rm{gas}}=0.039$~M$_{\odot}$, which is still above the total mass in the Solar System's planets ($0.002$~M$_{\odot}$) but slightly less than the classical minimum mass solar nebula (MMSN), derived from integrating an assumed gas surface density profile, $M_{\rm{MMSN}}=0.01$~M$_{\odot}$ \cite{weidenschilling1977}.  If the disk is more radially compact, e.g., $R\sim30-50$~AU \cite{hersant2001,adams2010,kretke2012}, the disk integrated \doh\ ratio would be lower as a direct consequence of removing mass where the gas is coldest.  In this instance, disk chemical processes, as determined by our model, could provide up to $<1\%$ of \doh\ in VSMOW for an outer disk radius of $R=50$~AU.  

We emphasize that we have not attempted to recreate a classical MMSN model but have opted instead to utilize a generic model based upon observations of protoplanetary disks in situ.  Two important points regarding the model are as follows.  First, the choice of a dust-settled model, chosen based upon the observational inference of a reduction of small grains from the upper layers of planet-forming disks \cite{furlan2006}, results in the disk being {\em{more}} permeable to X-rays, which are scattered by gas and dust, and thus the model adopted at present is {\em more favorable} towards deuterium fractionation than a well-mixed (uniform gas-to-dust ratio) disk model.  Second, because the mass of the disk is slightly lower than the typical MMSN, the average gas density would also be slightly lower than a MMSN model.  Thus ion-recombination is proportionally less efficient in a lower mass model, making for a higher steady state abundance of ions present for fractionation.  Both of these effects would lead to our fiducial model predicting higher \doh\ ratios than expected, and therefore our \doh\ results are likely to be upper limits, i.e., the total fractionation of water in the disk may be lower, and consequently requiring more water to be interstellar in nature.  To test the latter scenario, we re-ran all of our chemical models (described in detail below) with a factor of ten reduction in density and all other parameters held constant.  There was no significant change in the abundances relative to hydrogen calculated for the low-density model and so the net effect of the disk mass on the chemical results is small. 

The dust temperatures are calculated using the radiative transfer code TORUS \cite{harries2000,harries2004,kurosawa2004,pinte2009,cleeves2013a} where we have assumed stellar parameters typical of the Sun at $1-3$~Myr as described in the main text.  The gas temperatures are computed using a fitting function relating the local strength of the FUV field and disk gas density \cite{bruderer2012}.  The stellar FUV and X-ray radiative transfer is calculated using a 2D Monte Carlo treatment \cite{bethell2011u,bethell2011x}.  For the FUV radiation field we include absorption and scattering from dust in the continuum calculation, as well as resonant scattering off atomic hydrogen for Lyman-$\alpha$ propagation \cite{bethell2011u}. The X-ray treatment includes both photo-absorption and scattering by dust grains and gas \cite{bethell2011x}, which is essential as scattered X-ray photons provide the baseline midplane ionization in the absence of cosmic rays and radionuclide decay.  The physical model and radiation fields are shown in Fig.~S4.

\noindent \begin{center} {\bf Non-Stellar Ionization Processes} \end{center}

In addition to X-ray photoionization, our model includes non-stellar ionization from external galactic cosmic rays (GCR) and internal short-lived radionuclide decay.  In the Solar System, the solar wind is observed to efficiently modulate the GCR flux within a region known as the Heliosphere, especially at low GCR energies.  The net result is over a factor of $\gtrsim10$ reduction in the GCR ionization rate under {\em{present day}} solar minimum conditions.  T Tauri stars likewise drive stellar and disk winds, and thus it would not be unexpected for this mechanism to operate at equal or perhaps higher efficiency in the circumstellar environment of a T Tauri star.  The GCR ionization rate present in a disk has not yet been directly measured, but limits on the GCR ionization rate from H$_2$D$^+$ non-detections indicate that the GCR rate is sub-interstellar \cite{chapillon2011}.  Because the ionization rate is unknown in the young circumstellar environment, we  assume an unattenuated GCR ionization rate of $\zeta_{\rm{GCR}}=2\times10^{-19}$~s$^{-1}$, commensurate with the GCR rate under modern day solar maximum conditions \cite{cleeves2013a}.  From the specific solar maximum GCR spectrum we self-consistently vertically propagate GCR protons, including energy dependent losses through the gas disk, thus providing a comprehensive treatment of GCR ionization under the influence of wind modulation.  We emphasize that the GCR rate may be lower if winds are even more efficient at excluding cosmic rays from the natal environment, and thus deuterium fractionation powered by GCR ionization may be further hindered.    

Within the disk, near the geometrical midplane, contributions from the decay of radioactive particles become another important source of ionization, especially under the instance of GCR modulation.  The decay products, primarily $\gamma$-rays, positrons and electrons, originate from the decay of $^{26}$Al, $^{36}$Cl and $^{60}$Fe embedded in the refractory (dust) component.  For pebbles less than $a_{gr}\le1$~cm in size, the decay products can escape the pre-planetesimal prior to losing all of there energy \cite{umebayashi2009}. In the present model, we have included grain-growth up to $a_{gr}=1$~mm in the dust opacities, and thus we assume that all decay products escape the dust particles and are available for ionization, and the SLR abundances are taken to be the initial abundances present in the protosolar nebula \cite{cleeves2013b}.  Depending on location in the disk and the type of decay product considered, particles can be lost from the gas disk prior to losing all energy to ionization.  For example, $^{26}$Al $\beta$-decay results in the emission of a positron ($\Sigma\sim0.1$~g cm$^{-2}$) and energetic MeV photons ($\Sigma\sim12$~g cm$^{-2}$), which are each trapped at different disk surface densities and thus different radial locations in the disk.  Thus treatment of the losses of the radionuclide decay products is important in a complete picture of disk ionization, especially in the outer ($R>30$~AU) disk.  Furthermore, because our disk is settled, i.e., the dust and gas are no longer uniformly distributed, we calculate the vertical and radial position-dependent decay product losses, treating each decay product individually with regards to the relevant energy loss mechanism \cite{cleeves2013b}.  We note that if decay products are either trapped within dust particles or if the abundance of SLRs is lower than the protosolar value, the corresponding ionization rate due SLR decay would also be lower, further hindering D-fractionation. In Fig.~2 we show both the standard, fully present GCR ionization structure (left) next to our fiducial ionization model adopted in the main text (right), incorporating a mildly wind-excluded incident GCR ionization rate and detailed radionuclide transfer through the gas and dust disk, as described above.  

\noindent \begin{center} {\bf Chemical Reaction Network} \end{center}

We calculate time-dependent chemical abundances as a function of position throughout the disk utilizing a comprehensive disk chemistry code \cite{fogel2011,cleeves2013a}.  The backbone of the chemical network is based upon the Ohio State University's gas-phase reaction network \cite{smith2004}, which has been substantially expanded to include a host of chemical processes important in disks, including photodissociation, freeze-out, thermal and non-thermal sublimation, CO and H$_2$/HD/D$_2$ self-shielding \cite{wolcottgreen2011}, stellar and non-stellar ionization of H$_2$ and Helium \cite{fogel2011}.  In addition to the standard set, which encompasses 5912 reactions and 550 unique species, we have further expanded the network for the present study to include the essential deuterium fractionation reactions surrounding \htd\ chemistry, along with HDO. We have also expanded the set to include simple, primarily hydrogenation-based grain surface chemistry \cite{hhl}, which forms HDO/H$_2$O, HDCO/H$_2$CO, H$_2$, HD and D$_2$. It has been shown that deuterium enrichments are efficient in the gas phase \cite{millar1989} and in ices on dust grains \cite{tielens1983}. An additional fifty species are included in the expanded network, and the full set of reactions totals 6268.  The initial abundances in our model relative to total number of hydrogen atoms are listed in Table~S1.

Unless otherwise specified, the majority of the deuterated isotopologue reactions mirror those for the main isotopologues, where we have assumed the same reaction rate coefficients for both, and statistical branching ratios where appropriate.  In addition to H$_2$, HD and D$_2$ are directly ionized by X-rays, GCRs and SLRs.  Specific reaction rates are taken from the literature for the deuterated versions of the following reaction types:  (i) ${\rm{H_3^{+}+H_2}}$, ${\rm{H_3^{+}}}$ electron recombination and ${\rm{H_3^{+}+H}}$ \cite{roberts2004}; (ii) reactions with H$^+$ and atomic/molecular hydrogen, reactions with deuterated HCO$^+$, N$_2$H$^+$ and atomic hydrogen \cite{roberts2000}; and (iii) neutral-neutral warm deuterium reactions with barriers \cite{thi2010}.

There is one important exception in reaction set (i) regarding ${\rm{H_2D^{+}+H_2}}$, where we take into account the ortho- and para- spins of both reactants\cite{hugo2009}.  The ortho- and para- forms are not treated as distinct species but we instead assume that they are present with a thermal abundance ratio following\cite{cleeves2014}.  Including the spin information is important, as the energy barrier depends strongly on spin, with the reaction being barrierless if both reactants are ortho-type and $\Delta{E_1}=226$~K if both are para-type, for example.  Thus we calculate the local energy barrier for this reaction as a weighted mean of the ortho/para types in our determination of the reaction rate.  Finally, inclusion of the warm fractionation reactions in (iii) did not change our results.

To keep the network relatively chemically simple and transparent, we set out to form only singly deuterated water, HDO.
In the gas phase, deuterium enrichments are driven by ion-neutral reactions with \htd, where \htd\ reacts with oxygen, eventually leading to H$_2$DO$^+$, which can recombine with charged grains and electrons $\sim25\%$ of the time to form water (the rest of the time it goes to OH and either H or H$_2$).  On the grain surfaces, HDO forms by subsequent hydrogenation of oxygen ice \cite{tielens1982,hhl}.  Oxygen does not need to be ``permanently'' bound to the surface but simply needs to stay on the surface long enough for a hydrogen to adhere to a grain, sweep out the grain surface, and react with the heavier molecule (in this case, oxygen), which remains largely stationary.

\noindent \begin{center} {\bf Binding Energies} \end{center}

For most species, we assume the same  binding energies for deuterated and main isotopologues.  One exception is that we assume the surface binding energy of physisorbed deuterium is slightly higher (21~K) than hydrogen, $E_b(D)=471$~K and $E_b(H)=450$~K \cite{caselli2002}.  The distinction of {\em physisorbed} hydrogen stems from the observed fact that hydrogen has two main types of bonds on a substrate.  Physisorbed hydrogen weakly adheres to a surface by van der Waals forces with binding energies of around $E_b(H)=450$~K.  The second type is a far stronger chemical bond, called chemisorption, and has binding energies typically of order a few eV.  Formally each of these bonds corresponds to a different state of hydrogen on the grain surface, but because we do not track individual atoms we approximate this behavior by adopting a significantly higher chemisorption binding energy ($E_b=3000$~K) with respect to thermal and non-thermal desorption processes but adopt the physisorption energy when calculating the rate of hydrogen grain surface reactions, which proceed through the more mobile atoms. 

Another grain-surface facet of this study is a new treatment of the CO binding energy. CO is the second most abundance molecular volatile after H$_2$, and has been well studied in its binding properties on various substrates \cite{oberg2005,collings2004,bergin1995}.  The specific binding energy of CO to a substrate depends on the surface composition, namely if it is coated in water, CO, or CO$_2$ ices, or bare.  The CO on CO binding energy was determined to be $E_b=855$~K \cite{oberg2005}, corresponding to a dust temperature of $T_d\sim17$~K.  However, in regions of the disk close to and above this temperature, the surface seen by freshly adsorbed CO molecules will be predominantly non-CO by construction.  Again, because we do not track individual ice mantles nor their multi-layered structure, we approximate the binding surface based upon the local temperature.  At  dust temperatures below $T_d<25$~K, the ice will be primarily CO-dominated, and thus we adopt a binding energy of $E_b=855$~K.  Between $T_d=25-50$~K, approximately corresponding to temperatures between the freeze-out temperature of CO$_2$ down to that of CO, the ice mantle will be primarily CO$_2$.  CO$_2$ is a gas- and grain-surface product that readily proceeds in our chemical models in the presence of free oxygen or hydroxyl radicals.  At temperatures exceeding the freeze-out temperature of CO$_2$, $T_d>50$~K, the mantle will be primarily water ice, and above the water ice freeze-out temperature of $T\sim100$~K, the grains will be mostly bare.  The increased binding energy for CO on H$_2$O has been previously recognized \cite{collings2004,bergin1995}, however the consideration of CO$_2$ on CO is a new facet in the present study.

The relative binding energies of pure CO, CO on CO$_2$ ice and CO on H$_2$O ice were explored using temperature programmed desorption experiments of pure CO ice and thin CO layers on top of CO$_2$ and H$_2$O ices. Previous experiments have demonstrated that CO-CO and CO-H$_2$O binding energies differ substantially \cite{collings2003}, but less is known about the CO-CO$_2$ interaction. 

The experiments were carried out in a new ultrahigh vacuum (UHV) chamber (custom-made, Pfeiffer Vacuum), evacuated by a Pfeiffer Turbo HiPace 400 pump backed by a DUO 10M rotary vane pump to a base pressure of $\sim10$~mbar at room temperature. Ices are grown on a 2 mm thick IR transparent CsI substrate mounted on an optical ring sample holder through. The sample holder is connected to the cold tip of a closed cycle He cryostat (Model CS204B, Advanced Research Systems, Inc.) capable of cooling the CsI substrate down to 11 K. The cryostat is mounted on the top port of the chamber via a differentially pumped UHV rotary seal (Thermionics RNN-400) that allows 360 degree rotation of the CsI substrate inside the chamber without breaking the vacuum during the experiment. The CsI substrate is mounted onto the nickel-plated OHFC copper sample holder using silver gaskets for good thermal contact. A 50 ohm thermofoil heater is connected to the cryocooler tip so that the temperature of the substrate can be varied between $12-350$~K. The temperature of the substrate is controlled and monitored by a cryogenic temperature controller (Lake Shore Model 335) using two calibrated silicon diode sensors (accuracy of $\pm0.1$~K), one connected directly to the sample holder and the other near the heater element.  During heating, the temperature is increased using a linear heating ramp controlled by a positive feedback loop set using the Lake Shore 335 temperature controller.

Ices were grown onto the sample window using vapor deposition along the surface normal using de-ionized and freeze-thaw purified water, and high-purity CO and CO$_2$ gas ($>99\%$ purity guaranteed, which was confirmed by mass spectrometric measurements). The individual ice layer thicknesses were estimated from the deposition time and pressure and then quantified using transmission infrared spectroscopy using a Vertex 70v spectrometer with a liquid nitrogen cooled MCT detector and literature absorption coefficients \cite{gerakines1996}. In the presented experiments the pure CO ice is 16 monolayers (ML), the layered CO/CO$_2$ ice is 2/51 ML and the layered CO/H$_2$O ice is 3/46 ML. That is, in each experiment the ice thickness is sufficient to ensure that the underlying substrate is not affecting the desorption energies, and the top CO layers are thin enough that interactions with the underlying CO$_2$ and H$_2$O ices should dominate the desorption temperature of CO.

Following deposition, the ices were heated up using a linear 1~K/min heating ramp and the desorption rate of CO was monitored using a Pfeiffer quadrupole mass spectrometer (QMG 220M1, mass range 1-100 amu) positioned 40 mm of the CsI substrate. The resulting Temperature Programmed Desorption spectra are shown in Fig.~S5. Qualitatively, the desorption peaks of pure CO, CO$_2$ and H$_2$O are clearly separated. The desorption rate of CO peaks at ~28.5 K, the CO/CO$_2$ rate at ~37 K, and the CO/H$_2$O at ~44 K. The pure CO and CO/H$_2$O desorption behavior is consistent with previous studies \cite{oberg2005,collings2003}. The binding energy in Kelvin can be estimated from the peak position \cite{attard1998}, by multiplying the peak desorption rate temperature by 30. This yields binding energies of 855, 1110, and 1320 K for CO-CO, CO-CO$_2$ and CO-H$_2$O, respectively. Observe that this is a rather crude estimate and should mainly be used to constrain the relative binding energies of these three systems, while more experiments and more detailed modeling is required to set the absolute scale.

\noindent \begin{center} {\bf Chemical Model Tests} \end{center}
In our use of simplified deuterium network, there is concern that we may miss essential low-lying reactions that contribute over the duration of the chemical calculations or act in combination with a handful of other slow reactions that may impact our results.  To test this hypothesis, we have extracted the physical parameters (Table~S2) at a representative point at the outer disk midplane ($R=40$~AU) and the inner disk midplane ($R=3$~AU) from our disk model and have calculated the corresponding abundances using a more advanced general deuterium chemical network \cite{fduphd}.  The \doh\ values from our calculations (A) and the point-wise large scale network (B) are shown in Table~S3, where both models start with the same set of initial conditions (Table~S1).  For the most part the models are in good agreement, especially regarding the low \doh\ predicted for water, though some differences do exist.  The disparity in \htd/H$_3^+$ is a consequence of different assumptions for the reaction rates, where we have included the ortho-to-para spin information in the calculation of the rate coefficients \cite{hugo2009}.  In the B-model, $E_b=124$~K is assumed at both points.  The \htd/H$_3^+$ ratio in both cases is not directly imprinted into the water as a consequence of the oxygen being trapped in water, CO, and CO$_2$ ices and not available for new water chemistry, and thus the conclusion of little deuteration of water holds.  

\noindent \begin{center} {\bf Further Considerations} \end{center}
In the present section we compare how our choice of model parameters impacts the primary results of this paper.  To compare the main paper results to a model with GCRs fully present (unmodulated), we have recomputed our chemical models assuming a typical GCR rate (model type W98; \cite{cleeves2013a}).  The results are shown in Fig.~S6.  The HDO/H$_2$O ratio in ices is elevated from the main paper result. The layer of water ice where the \doh\ ratio is enriched is larger but remains centered on the warm molecular layer.  The maximum vertically integrated column of HDO/H$_2$O in ices approaches HDO/H$_2$O $\sim 10^{-4}$ at the disk outer edge ($R\sim80$~AU), which is still below VSMOW $\left({\rm{HDO/H_2O\sim3\times10^{-4}=2\times\left[D/H\right]_{VSMOW}}}\right)$. Recent models which include mixing have found that under a fully present GCR ionization rate, the \doh\ in water approaches $\rm \left[D/H\right]_{H_2O}=2\times10^{-3}$ after 1~Myr regardless of initial conditions and thus would allow for the \doh\ ratio to be mixed up from the molecular hydrogen value \cite{furuya2013}.  Because we do not include mixing we cannot compare with this claim directly, but in the instance of a low GCR ionization rate the efficacy of this process will be significantly curtailed.  In their models, oxygen is transported into the deuterium rich midplane to reform water.  The timescales for this process in the gas phase are too long compared to the freeze out time for oxygen onto grain surfaces regardless of ionization rate, so water reformation would have to be grain-surface chemistry driven. However, given the low abundance of hydrogen atoms to fuel hydrogenation ($\chi[{\rm H}]\sim10^{-13}$ and $\chi[{\rm D}]\sim10^{-17}$), grain surface reformation of water would not be efficient, and it is instead more favorable for the oxygen (or OH) to react with CO ices abundantly present on the grains to form CO$_2$ prior to getting doubly hydrogenated to make water, even though to CO reactions have a weak barrier, $E_b\sim80$~K \cite{garrod2011}.

Another component of our study was that we initiated the chemistry with all oxygen as gas-phase CO and water ice (Table~S1).  One could alternatively begin with nearly all oxygen in atomic form, i.e., fully available for chemical processing.  We computed a second model assuming the initial abundance of oxygen was $\chi({\rm{O}})=2\times10^{-4}$ and $\rm{\chi(H_2O~ice)=5\times10^{-5}}$ with HDO ice scaled appropriately to the protostellar molecular hydrogen \doh\ value.  The results are shown in Fig.~S7.  As can be seen, the HDO/H$_2$O ratio is slightly higher but still well below VSMOW, and thus our results regarding \doh\ in water do not depend strongly on where the oxygen is initially.  Furthermore, the model that begins with a substantial fraction of oxygen in atomic form rather than water only increases its water ice abundance by 0.2\% at 30~AU in the midplane after 1~Myr.  The lack of ionization not only hinders the formation of deuterated water (whose reactants are less abundant an take longer chemical times), but also the formation of the main isotopologue, H$_2$O, at the midplane.  Based on the knowledge that comets are comprised of a significant fraction of water by mass, approximately $\sim30\%$ \cite{greenberg1999} where silicates and refractory carbonaceous material comprise the rest, it seems unlikely that the disk formed from an initially primarily oxygen state, otherwise comets would be CO$_2$ ``snowballs'' rather than water ``snowballs.''  Above the midplane, where $z/r\gtrsim0.15$, the disk is able to convert the oxygen to water where there is sufficient ionization.  Nonetheless, the water formed in this gas still has just $\rm \left[D/H\right]_{H_2O}=5.3\times10^{-5}$ after 1~Myr, more than a factor of 3 below VSMOW.

Thus we find that regardless of initial oxygen abundances, chemical model assumed, and density structure, it is very difficult for disk chemistry alone to produce a significant amount of deuterated water ice (or perhaps any water ice) in the bulk gas at the outer disk, even at the VSMOW level, and thus the deuterium in water must have an interstellar heritage that has been since mixed down with warm deuterium-poor water to the values measured today.  Consequently, the survival of deuterium enrichments originating from interstellar ices will depend upon processing during the initial disk formation \cite{visser2009,yang2013} and the efficiency of mixing in the solar nebula between the active surface and the inert midplane, in addition to radial mixing with the warm inner ($\lesssim3$~AU) disk.

\newpage

\subsection{Supplementary Figures and Tables}

\renewcommand\thetable{{\bf S\arabic{table}}}

\begin{table}[h]
\caption{Log of the initial chemical abundances, $\chi$, per total number of hydrogen atoms. \label{tab:abun}}
\centering
\begin{tabular}{lr|lr} 
\centering
Species &  log($\chi$) & Species & log($\chi$)        \\               
\hline
H$_2$ & -0.30 & H$_2$O ice & -3.60 \\
HDO ice & -8.00 & He & -0.85 \\
N & -4.65 & CN & -7.22 \\
H$_3$$^+$ & -8.00 & CS & -8.40 \\
SO & -8.30 & Si$^+$ & -11.00 \\
S$^+$ & -11.00 & Mg$^+$ & -11.00 \\
Fe$^+$ & -11.00 & C$^+$ & -9.00 \\
CO & -4.00 & N$_2$ & -6.00 \\
C & -6.15 & NH$_3$ & -7.10 \\
HCN & -7.70 & HCO$^+$ & -8.05 \\
HD & -4.70 & H$_2$D$^+$ & -9.89 \\
HD$_2$$^+$ & -10.00 & D$_3$$^+$ & -9.70 \\
C$_2$H & -8.10 
\end{tabular}
\end{table}

\begin{table}[h]
\centering
\caption{Physical parameters for the chemical model comparison points in the midplane at the specified radii.
}
\begin{tabular}{l|ccccccr} 
\centering
& R  & n$_{\rm{H}}$ & T$_g$ & T$_d$ &$\Sigma~(\zeta_{\rm{H_2}})$  &$A^*_V$    \\           
 & (AU)  & (cm$^{-3}$) & (K) & (K) & (s$^{-1}$)     \\                  
\hline
 1 & 40 & $3.6\times10^{10}$  & 21.6 & 21.6& $1.1\times10^{-18}$ & $>20$  \\
2 & 3 &  $1.8\times10^{13}$ & 56.7  &56.7  & $1.5\times10^{-18}$ &  $>20$  
\end{tabular}
\end{table}

\begin{table}[h]
\centering
\caption{Independent chemical model comparison results.  Table values given with respect to \doh\ in H$_2$, i.e. $f[{\rm{R}}] = {\rm{R}}/4\times10^{-5}$.  A = chemical model presented in the main text; B = complex gas-grain deuterium chemical model \cite{fduphd}.}  
\begin{tabular}{c|ccc|ccc} 
\centering
 &    \multicolumn{3}{c|}{1} & \multicolumn{3}{c}{2}\\\hline
Ratio & A & B & A/B & A & B & A/B \\           
\hline
$f$[H$_2$D$^+$/H$_3^+$]            &  115      &  230   &    0.5 &  1.5      &  6.7 & 0.2\\
$f$[HDO/H$_2$O]                        &  1.0     &  1.1    &  0.9  &  1.0     & 1.1 & 0.9 \\
\end{tabular}
\end{table}

\newpage

\begin{table}[h]
\caption{\doh\ values and references from Fig.~1.}
\centering
\begin{tabular}{lccc}
\centering
Label & Value & Type & Reference  \\               
\hline
\hline
{\bf H$_2$} & & & \\
\hline
Protosun & $( 2.00 \pm  0.35)\times10^{-5}$ & Value & \cite{geiss2003} \\
Jupiter & $( 2.60 \pm  0.70)\times10^{-5}$ & Value & \cite{mahaffy1998} \\
Uranus & $( 4.40 \pm  0.40)\times10^{-5}$ & Value & \cite{feuchtgruber2013} \\
\hline
{\bf Water} & & & \\
\hline
Lunar Apatite & $( 1.12- 1.52)\times10^{-5}$ & Range & \cite{barnes2014} \\
Martian melt inclusions & $\le 1.99\times10^{-4}$ & Upper Limit & \cite{usui2012} \\
Semarkona Phyl. & $( 7.70 \pm  1.00)\times10^{-4}$ & Value & \cite{deloule1995} \\
CI Meteorites (Sub) & $( 6.43- 9.77)\times10^{-4}$ & Range & \cite{alexander2012} \\
CM Meteorites (Sub) & $( 8.29- 9.01)\times10^{-4}$ & Range & \cite{alexander2012} \\
Tagish Lake (Sub) & $( 5.00- 7.24)\times10^{-4}$ & Range & \cite{alexander2012} \\
Semarkona (Sub) & $( 2.80- 3.44)\times10^{-4}$ & Range & \cite{alexander2012} \\
J.F. Comet & $( 1.61 \pm  0.24)\times10^{-4}$ & Value & \cite{hartogh2011} \\
P/Halley & $( 3.02 \pm  0.07)\times10^{-4}$ & Value & \cite{eberhardt1995} \\
Hyakutake & $( 2.90 \pm  1.00)\times10^{-4}$ & Value & \cite{bockeleemorvan1998} \\
Hale-Bopp & $( 3.30 \pm  0.80)\times10^{-4}$ & Value & \cite{meier1998a} \\
Interstellar H$_2$O & $( 2.95- 9.50)\times10^{-4}$ & Range & \cite{persson2014} \\
IRAS 16293-2422 & $ 9.20^{+ 2.00}_{- 2.60}\times10^{-4}$ & Value & \cite{persson2012} \\
IRAS 16293-2422 Env. & $( 6.00 \pm  5.00)\times10^{-3}$ & Value & \cite{coutens2012} \\
Interstellar Ices & $\le 1.70\times10^{-2}$ & Upper Limit & \cite{parise2003} \\
\end{tabular}
\end{table}

\newpage

\setcounter{figure}{0}\renewcommand\thefigure{{\bf S\arabic{figure}}}

\begin{figure}[h]
\centering
\includegraphics[width=1.0\textwidth]{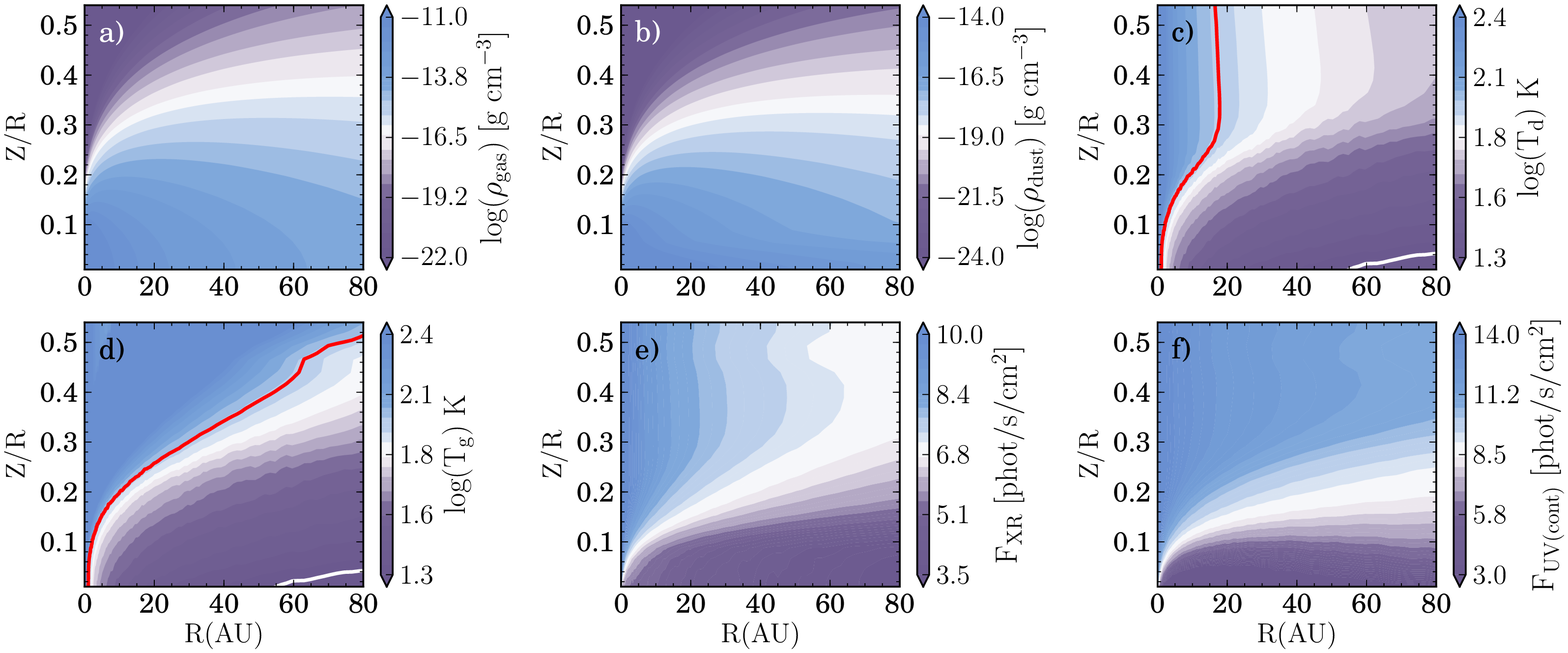}
\caption{Disk model assumed. Panels are: a) gas density, b) dust density, c) dust temperature, d) gas temperature, e) spectrally integrated X-ray flux, f) spectrally integrated continuum far UV flux.  In panels c) and d) red and white lines delineate $T=100$~K and $T=17$~K respectively.}
\end{figure}

\newpage

\begin{figure}[h]
\centering
\includegraphics[width=0.7\textwidth]{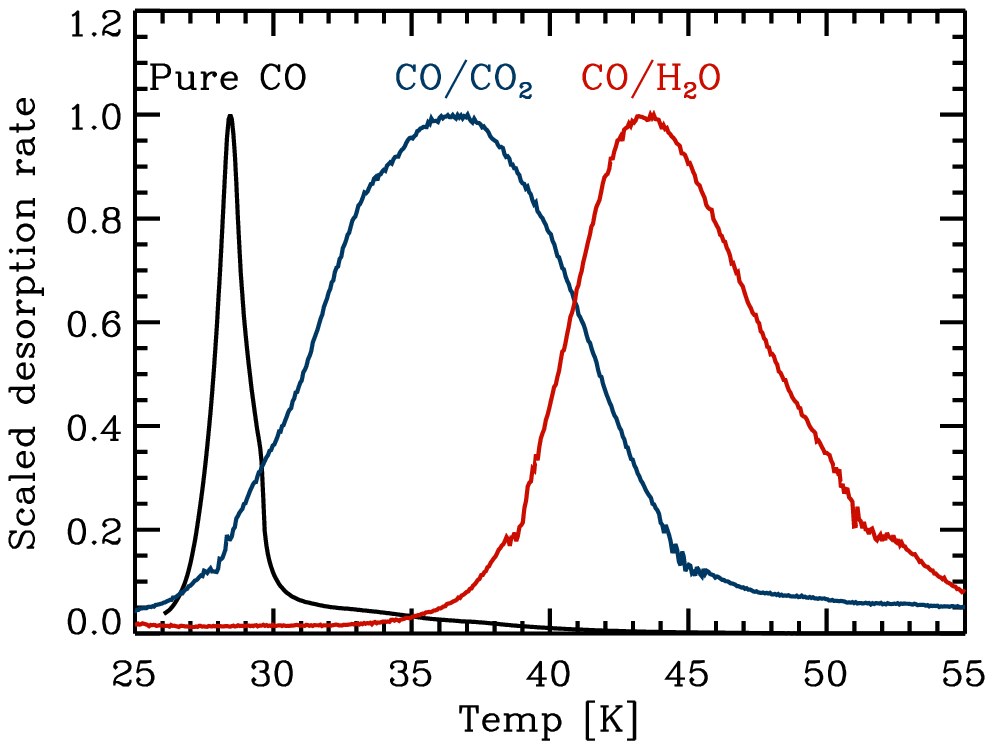}
\caption{Temperature programmed desorption spectra of pure CO, CO desorbing off a CO$_2$ ice, and CO desorbing of an amorphous H$_2$O ice. The desorption rates have been scaled to a unity peak desorption rate for visibility.}
\end{figure}

\newpage

\begin{figure}[h!]
\centering
\includegraphics[width=1.0\textwidth]{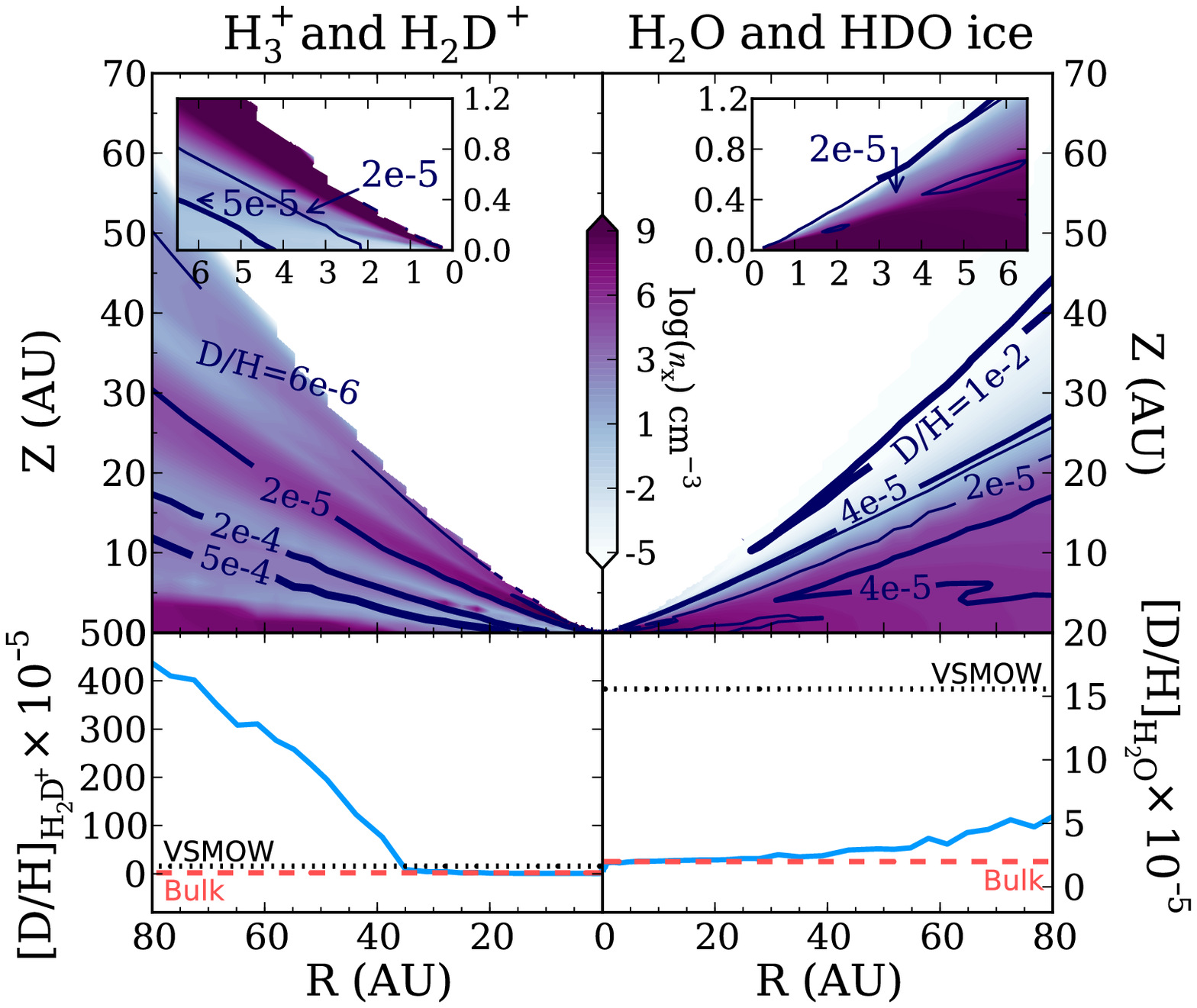}
\caption{Abundance and D-fractionation results for a standard ISM GCR rate.  The H$_3^+$ abundance is significantly elevated in the midplane compared to the fiducial model, and thus the vertically integrated \doh\ in \htd\ is substantially higher owing to the cold gas contribution to the column.  The \doh\ in water remains low, however, due to a lack of freely available oxygen for water reformation. Plot labels are the same as for Fig.~3.}
\end{figure}

\newpage

\begin{figure}[h!]
\centering
\includegraphics[width=1.0\textwidth]{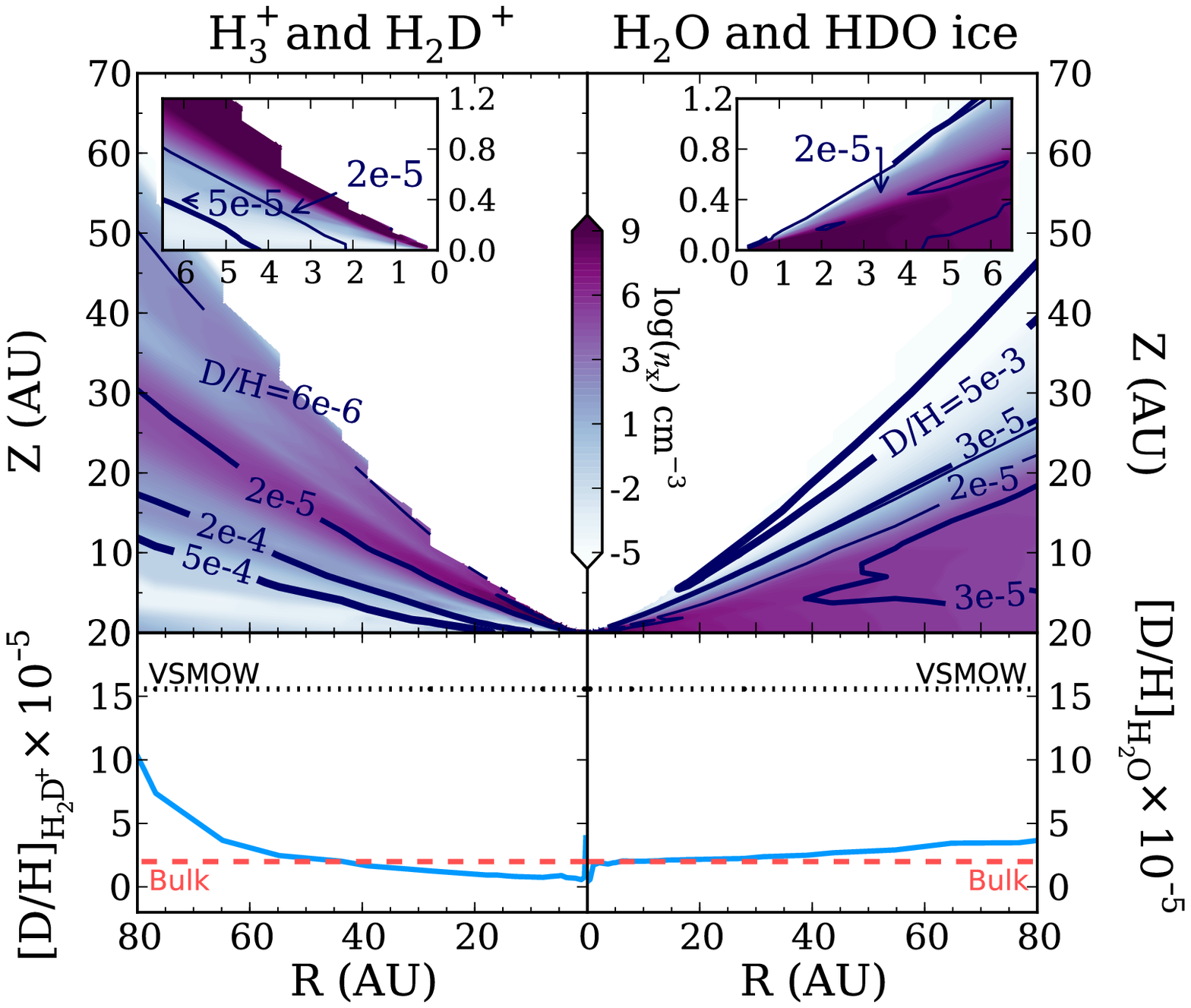}
\caption{Test case model where we began initially with most of the oxygen (that is not in CO) in gas-phase atomic O, rather than water ice.  Water \doh\ in ices remains low, even when we begin with atomic oxygen due to a lack of ionization. Thus the models are not strongly sensitive to the initial conditions of the oxygen. Plot labels are the same as for Fig.~3.}
\end{figure}

\end{document}